\documentclass[english,prl,twocolumn]{revtex4-1}
\usepackage[T1]{fontenc}
\usepackage{babel}
\usepackage{amsmath}
\usepackage{graphicx}

\begin{document}

\title{A comment on percolation and signatures of superconductivity in Au/Ag nanostructures}

\author{David Pekker, Jeremy Levy}
\affiliation{Department of Physics and Astronomy, University of Pittsburgh, Pittsburgh, PA 15260, USA \\
Pittsburgh Quantum Institute, Pittsburgh, PA  15260, USA
}

\date{\today}

\begin{abstract}
In this comment we point out that the experimental evidence for superconductivity presented by Thapa and Pandey in arXiv:1807.08572 is also consistent with a percolation transition. We propose simple follow-up experiments which would help to eliminate percolation as a possible explanation for the observed data. 
\end{abstract}

\maketitle

%Talking points:

Room-temperature superconductivity is a prospect that has captivated the imagination of scientists and engineers since the discovery of superconductivity in 1911 by Kammerling Onnes.
 The signatures of technological interest are simple enough to state: a zero-resistance state.  But there are other telltale features such as ``perfect'' diamagnetism (flux expulsion) that are expected from a superconductor.
 %(rather than just a ``perfect'' conductivity).

There has been a recent report by Thapa and Pandey~\cite{Thapa2018} claiming room-temperature (and above-room-temperature) superconductivity in gold/silver dispersed nanoparticles.  Four-terminal transport measurements are reported, showing a sharp transition to a zero-resistance state (within measurement limits) below a critical temperature.  Some lowering of the transition temperature with increasing applied magnetic fields is also reported.  Magnetization measurements show a transition to a state with large diamagnetism at a temperature that coincides with the resistive transition. Taken together, these results appear to provide compelling evidence for superconductivity at ``room'' temperature.  

It was subsequently noted by Skinner~\cite{Skinner2018} that there is an unexplained apparent duplication of ``noise'' in two of the diamagnetic curves shown in the manuscript.  There is no obvious physical explanation for why this noise should appear to precisely (up to an overall offset) repeat itself.

\begin{figure}[b]
\includegraphics[width=\columnwidth]{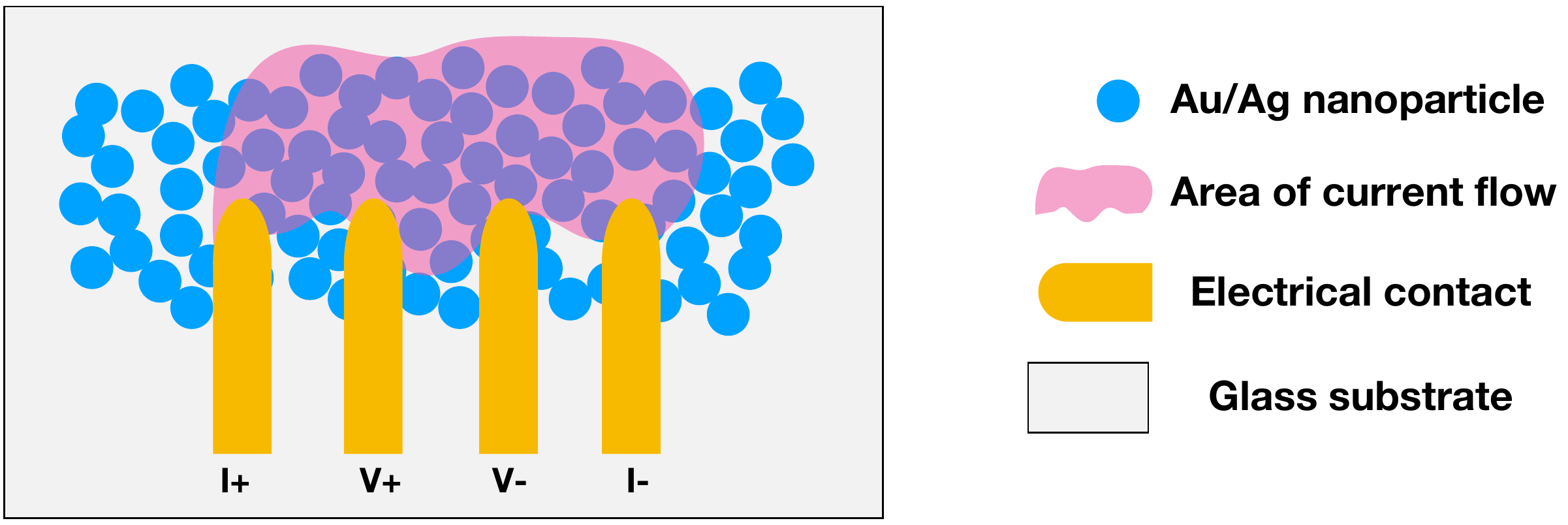}
\caption{Device layout}
\label{fig:layout}
\end{figure}

Here, we set aside the question of the origin of this unexplained noise correlation, and focus on the resistive transition itself.  The question we wish to address is whether there is an alternate physical explanation that could give rise to this zero-resistance signature.  Below, we describe one such scenario based on a temperature-driven percolation transition and suggest simple follow-up experiments that could rule out this non-superconducting interpretation.

Figure 1 schematically depicts the device layout. A collection of gold/silver nanoparticles are dispersed onto a glass slide, on top of an array of electrical contacts.  Electrical measurements are conducted by applying a known current through leads $I_+$ and $I_-$, and measuring the voltage difference $V_+ - V_-$.  (The precise layout of the voltage and current leads are not explicitly specified in the manuscript, but are presumed to be aligned as shown in Figure 1.)  The region of the sample where current is flowing is represented by the pink shading.

\begin{figure}[b]
\includegraphics[width=0.8\columnwidth]{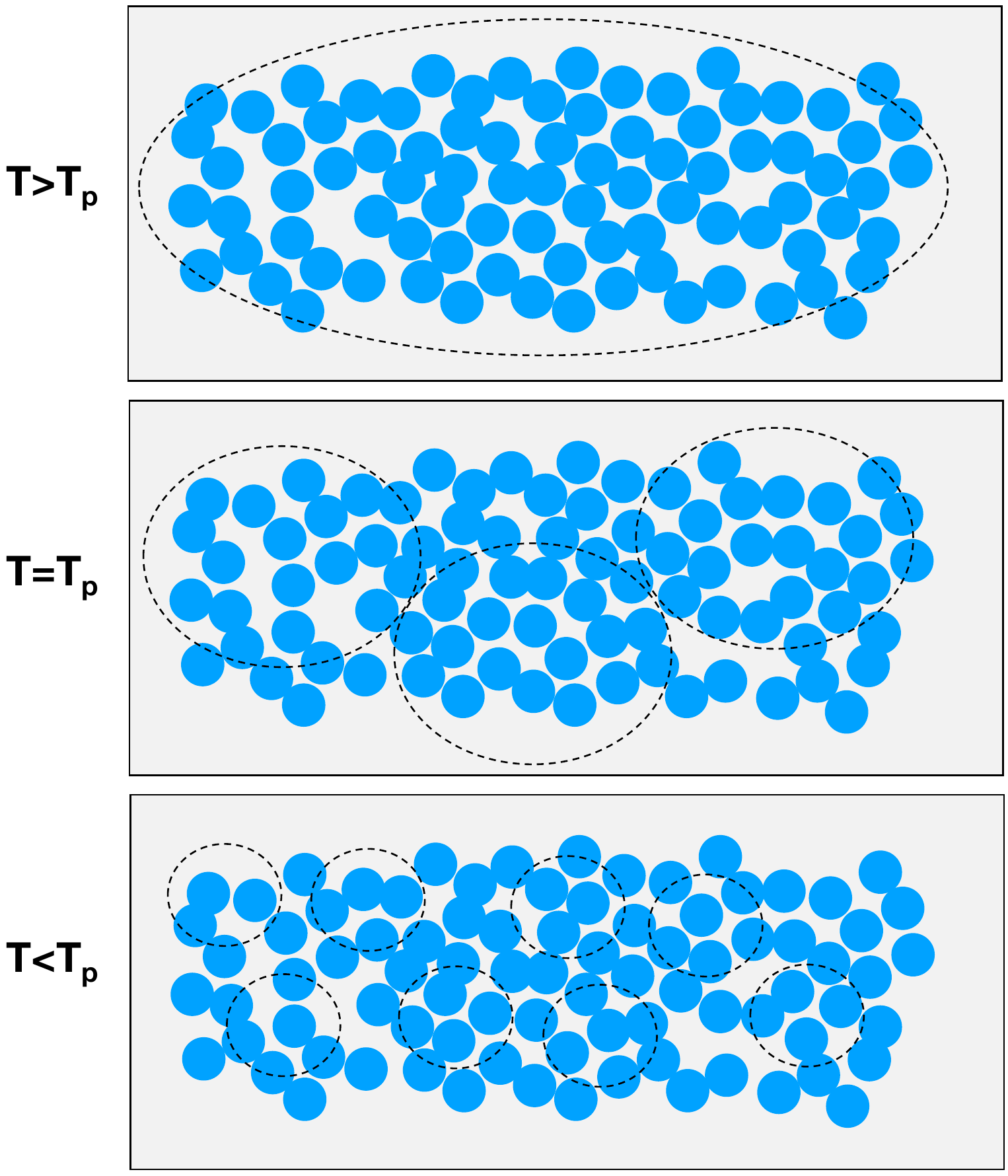}
\caption{Schematic showing conducting domains across the temperature driven percolation transition.}
\end{figure}

Let us consider the possibility that there is a percolation transition (Figure 2) at $T=T_p$ such that for $T>T_p$ the entire film is continuously connected, while for $T<T_p$ the film breaks up into disconnected regions. Percolation transitions have been extensively studied in the past, see e.g.~\cite{Stauffer1994}. Additionally, there are previous reports~\cite{Doty2001} of such percolative transitions in material systems that bear a striking resemblance to the Thapa and Pandey~\cite{Thapa2018} report.  

We now argue that a percolative transition can plausibly lead to an apparent zero-resistance state. Figure 3(a) depicts a configuration below $T_p$ in which conducting domains are disjoint (depicted by dotted ellipses). Upon the application of a sufficiently high voltage bias across leads 1 and 4, a conduction path is formed between these two leads. In this scenario, the current passes between the outer leads and bypasses the inner voltage leads completely.  As a consequence, the two voltage leads can either be connected to one another or float to the same potential, resulting in a zero-voltage difference $V_{32}$ between them (see Figure 3(b)).  As the current $I_{14}$ is increased further, the voltage $V_{32}$ could be expected to rise abruptly above some higher breakdown current.

\begin{figure}[t]
\includegraphics[width=\columnwidth]{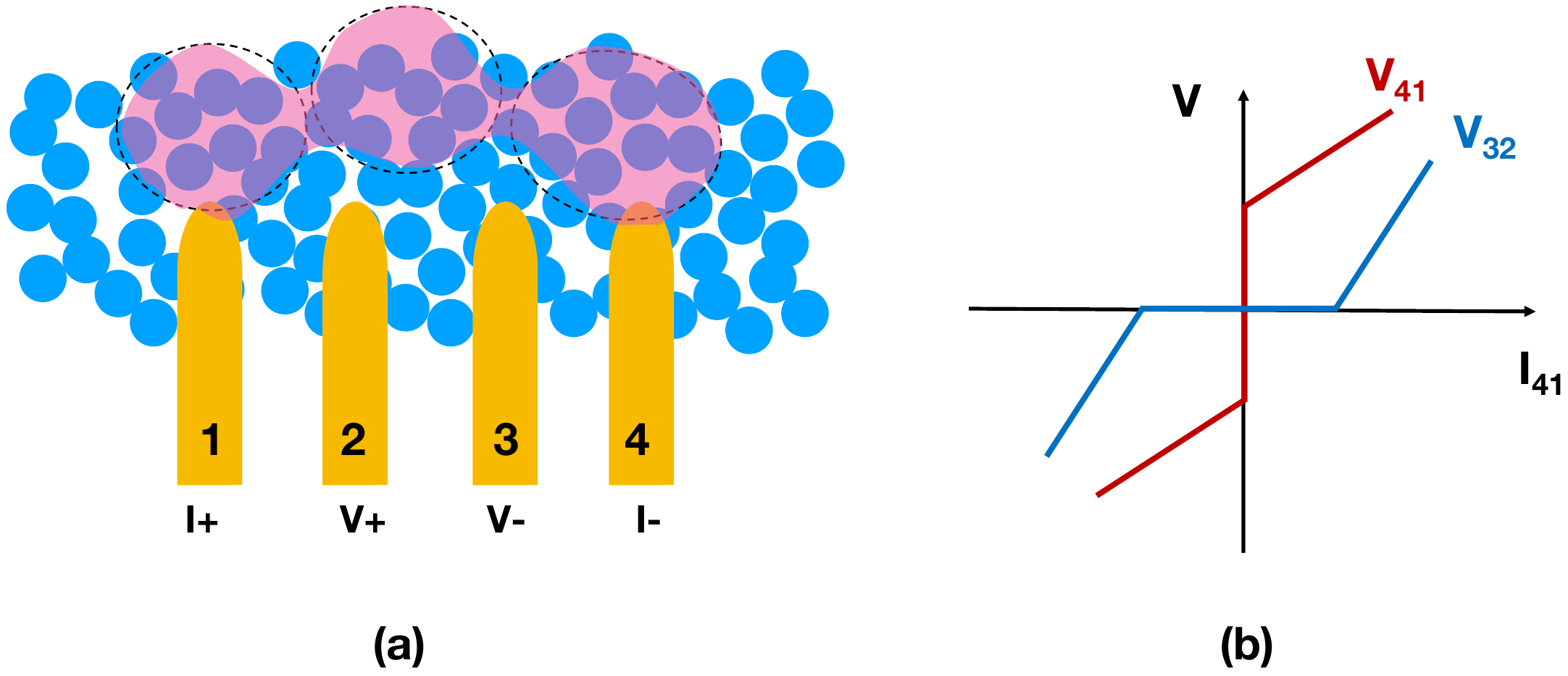}
\caption{Alternative scenario: below the percolation threshold a current carrying state could appear to have zero resistance in a four-terminal measurement.}
\end{figure}

%The samples described are an agglomeration of gold/silver nanoparticles that are separated by distances that suggest a proximity to a percolation transition.  Near this transition, the number of conduction pathways connecting one end of the device to another begins to approach zero, in ways that have been extensively investigated previously.  If indeed the sample is close to such a transition, it would be sensitive to very small changes in separation between adjacent nanoparticles, and differential thermal expansion could potentially lead to local changes in conducting pathways.  Figure 1(right) illustrates a situation where at high temperatures, the conducting path in a four-terminal measurement from a configuration in which the current path overlaps both the current leads (source, drain) and voltage probes.  At lower temperature, if the current path were to exclude the voltage leads, then it is possible that the voltage leads may be shorted to one another but disconnected from the current path.  In that situation, the apparent voltage drop across the sample would be zero, and the calculated resistance would also be zero.

The question arises as to how such a transition could take place and depend on temperature.  Again, the percolation condition may be very sensitive to minute changes in separations between the nanoparticles.  Further, alloying could change the temperature at which this transition occurs.  Glass, gold, and silver all have different thermal expansion coefficients, from $\sim 9\times 10^{-6}$ K$^{-1}$  for glass, to $14 \times 10^{-6} \rm{K}^{-1}$ for gold, and $18 \times 10^{-6} \rm{K}^{-1}$  for silver. The differential thermal contraction with the substrate could easily drive the percolations transition, see e.g.~\cite{Doty2001}. 

The second major piece of evidence presented relates to the observed onset of diamagnetism below the critical temperature.  Gold is a well-known diamagnetic material in bulk form~\cite{Suzuki2012}. There are reports that diamagnetism is strongly suppressed in thin films~\cite{Reich2006}. These observations indicate that  percolation could strongly effect the diamagnetic properties of the nanostructured material.

We conclude by pointing out that further transport measurements, not described in the original manuscript, could help to elucidate what is going on. Specifically, we suggest measuring a series of two-terminal I-V characteristics as a function of temperature. There should be a resistance reduction at the superconducting transition, even taking into account the measurement lead resistance.  A more sensitive probe of the scenario outlined above would involve measuring I-V curves between the current and voltage leads (e.g., between leads 1 and 2) to evaluate whether the voltage leads are electrically connected to the current paths between leads 1 and 4.  Additionally, other permutations of voltage and current leads could help elucidate whether there are fully connected paths between all of the measuring leads.

%If below the critical temperature these characteristics resemble the curve labeled $V_{41}$ in Figure 3(b), or if the two-terminal resistance increases below the critical temperature, then a percolation explanation is likely correct. On the other hand, if below the critical temperature the two terminal I-V characteristics resemble the curve labeled $V_{23}$ in Figure 2b, or the two terminal resistance decreases with temperature, than the percolation explanation can be eliminated. Additionally, we suggest permuting the voltage and current leads in the four-terminal measurement in order to eliminate the possibility that the voltage leads are not connect to the current path. 

% \begin{figure*}
% \includegraphics[width=14cm]{money}
% \caption{Illustration of how a percolation transition could be misinterpreted as evidence for high-conductivity.}
% \end{figure*}


\begin{thebibliography}{99}
\bibitem{Thapa2018} Dev Kumar Thapa and Anshu Pandey, arXiv:1807.08572.
\bibitem{Skinner2018} Brian Skinner, arXiv:1808:02929.
\bibitem{Stauffer1994} D. Stauffer and A. Aharony, {\it Introduction to Percolation Theory.\/} (Taylor \& Francis, London, 1994).
\bibitem{Doty2001} R. Christopher Doty, Hongbin Yu, C. Ken Shih, and Brian A. Korgel, J. Phys. Chem. B, {\bf 105}, 8291 (2001).
\bibitem{Suzuki2012} Motohiro Suzuki, Naomi Kawamura, Hayato Miyagawa, Jose S. Garitaonandia, Yoshiyuki Yamamoto, and Hidenobu Hori, PRL {\bf 108}, 047201 (2012).
\bibitem{Reich2006} S. Reich, G. Leitus, and Y. Feldman, APL {\bf 88}, 222502 (2006).
\end{thebibliography}
\end{document}